\documentstyle[prb,aps,epsf,preprint]{revtex}
\begin{document}
\draft

\title{A transferable nonorthogonal tight-binding model of germanium: application to
small clusters}

\author{Jijun Zhao $^*$}
\address{Department of Physics and Astronomy, University of North Carolina
at Chapel Hill, Chapel Hill, NC 27599, USA. \\
International Centre for Theoretical Physics, P.O.Box 586, Trieste 34100, Italy}

\author{Jinlan Wang, Guanghou Wang}
\address{National Laboratory of Solid State Microstructures, Nanjing
University, Nanjing 210093, P.R. China}

\date{\today}
\maketitle
\begin{abstract}

We have developed a transferable nonorthogonal tight-binding total energy model 
for germanium and use it to study small clusters. The cohesive energy, bulk modulus, 
elastic constants of bulk germanium can be described by this model to considerably 
good extent. The calculated bulk phase diagram for germanium agrees well with LDA 
results. The geometries and binding energies found for small Ge$_n$ clusters with 
$n=3-10$ are very close to previous {\em ab initio} calculations and experiments. 
All these results suggest that this model can be further applied to the
simulation of germanium cluster of larger size or with longer time scale,
for which {\em ab initio} methods is much more computational expensive.
\end{abstract}


In the past decade, tight-binding molecular dynamics (TBMD) has evolved
into a powerful approach in the simulation of semiconductor materials
\cite{1,2,3}. In the tight-binding scheme, although the system is still
described in a quantum-mechanical manner, the computational cost is
significantly reduced due to the parameterization of Hamiltonian matrix
elements. In many cases of material simulations, it might offer a satisfactory 
compromise between empirical \cite{4} and first principle \cite{5,6} methods
for modeling the interatomic interaction. As an alternative of the accurate
but costly {\em ab initio} molecular dynamics, TBMD can handle more
complicated systems with acceptable accuracy \cite{1,2,3}.

For carbon and silicon, there are several well established orthogonal 
\cite{7,8,9} and nonorthogonal \cite{10,11,12} tight-binding models. 
Although the orthogonal models works well for various bulk systems
\cite{1,2,3}, Menon found that the inclusion of the nonorthogonality 
of tight-binding basis is essential for describing the geometries 
and binding energies of small silicon clusters \cite{11,12}.
Compared to carbon and silicon, there is much fewer tight-binding models
developed for germanium. Recently, M.Menon has extended the nonorthogonal
tight-binding (NTB) scheme to germanium and calculated the structures and
cohesive energies of small Ge$_n$ clusters \cite{13}. Although the cluster
geometries obtained in Ref.[13] are generally consistent with {\em ab initio}
results, the binding energies are overestimated.
In this work, we perform an independent fitting of NTB parameters
for germanium, which describes binding energies of germanium clusters better 
than that in Ref.[13]. This model is employed to study some bulk properties 
and considerably good results are obtained. 

In Menon's NTB model \cite{11,12,13}, the total binding energy $E_{b}$ of 
a system with $N_a$ atoms can be written as a sum
\begin{equation}
E_{b}=E_{el}+E_{rep}+N_aE_{0}
\end{equation}
$E_{el}$ is the electronic band energy, defined as the sum of
one-electron energies $\epsilon _k$ for the occupied states: 
$E_{el}=\sum_k^{occ}\epsilon _k$. In Eq.(1), a constant energy correction
term $N_aE_0$ and a repulsive interaction $E_{rep}$ are also included.

On nonorthogonal basis set, the eigenvalues $\epsilon _k$ of system
are determined from the secular equation:
\begin{equation}
{\rm det}|H_{ij}-\epsilon S_{ij}|=0.
\end{equation}
Here the overlap matrix elements $S_{ij}$ are constructed in the spirit of
extended H\"uckel theory \cite{14},
\begin{equation}
S_{ij}=\frac{2V_{ij}}{K(\epsilon _i+\epsilon _j)}
\end{equation}
and the nonorthogonal Hamiltonian matrix elements by 
\begin{equation}
H_{ij}=V_{ij}[1+\frac {1}{K}-S_2^2]
\end{equation}
where 
\begin{equation}
S_2=\frac{(S_{ss\sigma }-2\sqrt{3}S_{sp\sigma }-3S_{pp\sigma }+3S_{pp\pi })}4
\end{equation}
is the nonorthogonality between two $sp^3$ bonding orbitals and $K$ is a
environment dependent empirical parameter \cite{11}.

The $H_{ij}$ and $S_{ij}$ depend on the interatomic distance through the 
universal  parameters $V_{ij}$, which are calculated within Slater-Koster's 
scheme \cite{15}. The scaling of the Slater-Koster parameters 
$V_{\lambda \lambda ^{\prime }\mu }$ is taken to be exponential with the
interatomic distance $r$
\begin{equation}
V_{\lambda \lambda ^{\prime }\mu }(r)=V_{\lambda \lambda ^{\prime }\mu
}(d_0)e^{-\alpha (r-d_0)}
\end{equation}
where $d_0=2.45 \AA$ is the bond length for germanium crystal in the diamond
structure \cite{16}.

The repulsive energy $E_{rep}$ in Eq.(1) is given by the summation of
pairwise potential function $\chi (r)$:
\begin{equation}
E_{rep}=\sum_i\sum_{j>i}\chi (r_{ij})=\sum_i\sum_{j>i}\chi _0e^{-4\alpha (r_{ij}-d_0)}
\end{equation}
where $r_{ij}$ is the separation between atom $i$ and $j$.

In practice, we adopt the Slater-Koster hopping integrals
$V_{\lambda \lambda ^{\prime }\mu }(d_0)$ fitted from the band structure
of bulk germanium \cite{17}. The on-site orbital energies $\epsilon_s$,
$\epsilon_p$ are taken from atomic calculations \cite{18}.
The only four adjustable parameters $\alpha$, $K$, $\chi _0$, $E_0$ are 
fitted to reproduce the fundamental properties of germanium bulk and dimer.
The input properties include: the experimental values of
bulk interatomic distance 2.45 $\AA$ \cite{16}, dissociation
energy 2.65 eV \cite{19} and vibrational frequency (286$\pm$5 cm$^{-1}$)
\cite{20} of Ge$_2$ dimer, as well as theoretical bond length of Ge$_2$ 
(2.375 \AA) from accurate quantum chemistry calculation at G2(MP2) 
level \cite{21}. The fitted parameters are given in Table I.

\begin{table}
\ \\ Table I. Parameters in the NTB mode developed for germanium in this work.
See text for detailed descriptions. 
\begin{center}
\begin{tabular}{cccccc}
$\epsilon_s$ & $\epsilon_p$ & $V_{ss\sigma}$ & $V_{sp\sigma}$ & $V_{pp\sigma}$
& $V_{pp\pi}$ \\ \hline
 -14.38 eV & -6.36 eV & -1.86 eV & 1.90 eV & 2.79 eV & -0.93 eV
\ \\
\end{tabular}
\ \\
\begin{tabular}{ccccc}
$d_0$ & $K$ & $\chi_0$ & $\alpha$ & $E_0$ \\ \hline
2.45 $\AA$ & 1.42 & 0.025 eV  & 1.748 $\AA^{-1}$ & 0.79 eV \\
\end{tabular}
\end{center}
\end{table}

We can first check the validity of the present NTB scheme by studying the
fundamental properties of germanium solid in diamond phase. The obtained
cohesive energy 3.58 eV/atom is very close to experimental value 3.85 eV/atom
\cite{16}. Furthermore, we have calculated the bulk modulus $B$ and
elastic constants $C_{11}$, C$_{12}$, C$_{44}$ of germanium and compared
with experimental values \cite{22} in Table II. Most of the bulk elastic 
properties such as $B$, $C_{11}$, C$_{12}$ are well reproduced 
except that the C$_{44}$ is overestimated by 0.35 Mbar in our model.

\begin{table}
\ \\
Table II. Bulk modulus and elastic constants (in units of Mbar) of bulk
germanium in diamond structure. NTB are the theoretical results from
present NTB model; Exper. denote the experimental values taken from Ref.[22].
\begin{center}
\begin{tabular}{ccccc}
        & C$_{11}$ & C$_{12}$ & C$_{44}$ &   $B$  \\ \hline
NTB     &  1.125   &  0.545   & 1.019    & 0.738  \\
Exper.  &  1.288   &  0.483   & 0.671    & 0.751  \\
\end{tabular}
\end{center}
\end{table}

\begin{figure}
\centerline{
\epsfxsize=3.0in \epsfbox{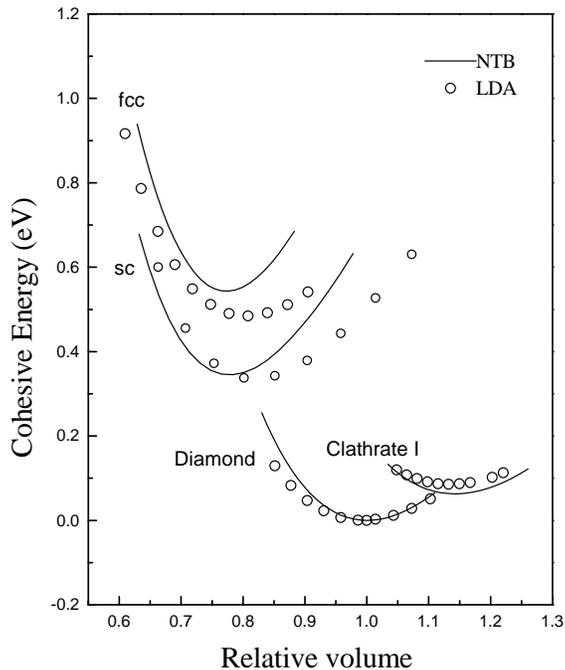}
}
\caption{Cohesive energies vs. relative volume for bulk germanium in simple cubic 
(sc), diamond and clathrate I phase from NTB (solid line) and LDA (open circle) 
calculations \cite{23}.}
\end{figure}

By using the NTB scheme, we have also calculated the equation of states of
germanium in different phases. In Fig.1, we present the zero-temperature
phase diagram of the fcc, sc, diamond and type I clathrate obtained
from NTB model, along with recent LDA plane-wave pseudopotential calculations
\cite{23}. It is worthy to noted that our NTB model is able to described the
energy and atomic volume of clathrate phase. The energy of clathrate I is 
0.06 eV/atom higher than that of diamond phase and its relative volume is 
about 15$\%$ larger than diamond phase. These results are consistent with 
the 0.08 eV energy difference and $15\%$ volume change from LDA calculation 
\cite{23}. The success in clathrate phase is important since the 
clathrate is also four-coordinated structure \cite{23}. On the other hand,
it is natural to find that the agreement between LDA and NTB scheme become 
worse in the high-coordinated phases like fcc since the present model is fitted 
from the diamond phase and dimer. However, the relative poor description of 
high coordinate phase will not influence the study on germanium clusters 
since such high coordination ($\sim$ 12) does not exist in the geometries 
of germanium clusters. Considering its extreme simplicity and small number 
of adjustable parameters, the current NTB scheme gives a sufficient satisfactory 
overall description of bulk germanium properties. Therefore, one can expect that 
the model to give a reasonable description on the germanium clusters. 

In this paper, we determine the lowest energy structures of the Ge$_n$ clusters 
with $n=3-10$ by using TBMD full relaxation. The ground state structures of Ge$_n$
($n=5-10$) are presented in Fig.2 and the geometrical parameters of small
Ge$_n$ ($n=3-7$) clusters are compared with previous {\em ab initio}
calculations \cite{24,25,26,27} in Table III. In general, both the lowest energy 
structures and their characteristic bond length agree well with 
{\em ab initio} results. A brief description is given in the following.

\begin{figure}
\centerline{
\epsfxsize=3.0in \epsfbox{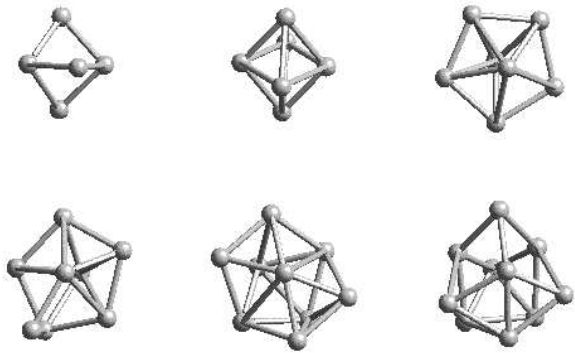}
}
\caption{Lowest energy structures of Ge$_n$ ($n=5-10$) clusters.}
\end{figure}

\begin{table}
\ \\
Table III. Lowest energy geometries (with characteristic bond length parameters)
of small Ge$_n$ clusters. Tight-binding calculation (NTB) are compared
with previous {\em ab initio} results such as: MRSDCI \cite{24}, B3LYP \cite{25},
LDA \cite{26,27}. The label of atom and bond for Ge$_n$ are taken from Ref.29.
\begin{center}
\begin{tabular}{cccccccc}
$n$& Sym.   &Bond& \multicolumn{5}{c}{Bond length (\AA)}  \\ \hline
   &         &    &MRSDCI\cite{24}&B3LYP\cite{25}&LDA\cite{26}&LDA\cite{27}& NTB\\ \hline
3  &$C_{2v}$ &1-2 & 3.084 & 3.070 &  3.20      & 2.91  & 2.71 \\
   &         &1-3 & 2.320 & 2.312 &  2.26      & 2.21  & 2.38 \\
4  &$D_{2h}$ &1-2 & 2.477 & 2.475 &  2.40      & 2.35  & 2.44 \\
   &         &1-3 & 2.622 & 2.619 &  2.53      & 2.44  & 2.57 \\
5  &$D_{3h}$ &1-2 & 3.277 & 3.135 &  3.19      & 3.10  & 2.87 \\
   &         &1-3 & 2.456 & 2.476 &  2.39      & 2.34  & 2.44 \\
   &         &3-4 & 2.456 & 3.320 &  3.19      & 3.10  & 3.40 \\
6  &$O_{h}$  &1-2 &  --   & 2.553 &  2.47      & 2.40  & 2.47 \\
   &         &2-3 &  --   & 2.941 &  2.85      & 2.78  & 2.70 \\
7  &$D_{5h}$ &1-2 &  --   &  --   &  2.65      & 2.56  & 2.83 \\
   &         &1-3 &  --   &  --   &  2.57      & 2.49  & 2.57 \\
   &         &3-4 &  --   &  --   &  2.59      & 2.51  & 2.53 \\
\end{tabular}
\end{center}
\end{table}

The minimum energy structure found for Ge$_3$ is an isosceles triangle
(C$_{\rm 2v}$) with bond length 2.38 \AA$~$ and apex angle $\theta=69.5^{\circ}$,
in agreement with {\em ab initio} calculations (see Table III for comparison). The
linear chain has higher total energy of about 0.95 eV. 

The ground state structure of germanium tetramer is a planar rhombus
(D$_{\rm 2h}$) with side length 2.44 \AA $~$ and minor diagonal length 2.57
\AA. This structure has been predicted as ground state in all {\em ab initio}
calculations \cite{21,24,25,26,27,28} and the tight-binding bond length are close 
to {\em ab initio} results.

For the Ge$_5$, the lowest energy configuration is obtained as a
strongly compressed trigonal bipyramid (D$_{\rm 3h}$). The energy of structure 
is lower than the perfect trigonal bipyramid by 0.62 eV and the planar edge capped 
rhombus by 0.35 eV. The trigonal bipyramid structure has been considered in all of 
the previous {\em ab initio} studies \cite{21,24,25,26,27,28}. In those LDA based 
simulation without symmetry constraint \cite{25,26,28}, the trigonal bipyramid is 
found to undergo severe compression and relax to the structure in Fig.2.

A distorted octahedron (D$_{\rm 4h}$) is obtained for Ge$_6$ as lowest energy
structure. This structure is found to be energetically degenerated with a
edge-capped trigonal bipyramid ($\Delta E=0.018$ eV). This result agree
well with recent B3LYP and Car-Parrinello calculation of Ge$_6$ \cite{27,28}.

In the case of Ge$_7$, we find a compressed pentagonal bipyramid with
D$_{\rm 5h}$ symmetry as ground state and energetically lower than the face
capped octahedron by 0.63 eV. The pentagonal bipyramid structure 
has also been obtained from LDA based simulations \cite{25,26,28}.

\begin{table}
\ \\ Table IV. Binding energy per atom $E_{b}/n$ (eV) of Ge$_n$ clusters
obtained within the present NTB model, compared to experimental values
\cite{19,30}, {\em ab initio} results based on 
G2(MP2) level\cite{21} or LDA plane-wave pseudopential \cite{26,28},
as well as nonorthogonal tight-binding \cite{13} calculations.
\begin{center}
\begin{tabular}{ccccccc}
$n$&Exper.\cite{19,30}&G2(MP2)\cite{21}&LDA\cite{26}&LDA\cite{27}&NTB\cite{13}&NTB(present)\\ \hline
2  &   1.32   & 1.25      & 1.89   &  --    & 1.31   & 1.32  \\
3  &   2.24   & 2.02      & 2.78   & 2.66   & 2.11   & 2.06  \\
4  &   2.60   & 2.49      & 3.32   & 3.19   & 2.66   & 2.62  \\
5  &   2.79   & 2.68      & 3.58   & 3.45   & 2.85   & 2.73  \\
6  &   2.98   &  --       & 3.76   & 3.63   & 3.05   & 2.95  \\
7  &   3.03   &  --       & 3.90   & 3.77   & 3.19   & 3.09  \\
8  &   3.04   &  --       & 3.82   & 3.69   & 3.17   & 3.05  \\
9  &   3.04   &  --       & 3.93   & 3.79   & 3.25   & 3.12  \\
10 &   3.13   &  --       & 4.04   & 3.91   & 3.32   & 3.17  \\
\end{tabular}
\end{center}
\end{table}

An additional atom capped to pentagonal bipyramid of Ge$_7$ yields the 
lowest energy structure for Ge$_8$. This structure
is more stable over the bicapped octahedron by 0.08 eV. Both of these two 
structures are found for Ge$_8$ in Car-Parrinello simulation, while
bicapped octahedron is lower in energy by 0.03 eV \cite{28}.

A bicapped pentagonal bipyramid is found for Ge$_9$. It is more stable than
a capped distorted square antiprism by 0.06 eV. The current ground state structure
has been found in Car-Parrinello simulation for Ge$_9$ \cite{28} but it is
0.08 eV higher than the capped square antiprism structure.

For Ge$_{10}$, the tetracapped trigonal prism (C$_{\rm 3v}$) is found to be
most stable and 0.16 eV lower than the bicapped square antiprism (D$_{\rm 4d}$).
This ground state structure is consistent with previous LDA results \cite{26,28}.

In Table IV, we compare the binding energy per atom $E_{b}/n$ 
for Ge$_n$ ($n=2-10$) with the other theoretical and available experimental results. 
Due to the local density approximation, LDA calculation \cite{26,28} has 
systematically overestimated the cluster binding energies. The more accurate 
binding energies for small germanium clusters up to five atoms has been provided 
by a more sophisticated G2(MP2) computation \cite{21}. Although all the empirical 
parameters in our NTB model are fitted from dimer and bulk solid and there is no 
bond counting correction included, the experimental cohesive energies are fairly 
well reproduced by our calculation. Typical discrepancy between our calculation 
and experiment is less than 0.1 eV for those clusters. The successful description
of binding energy within the present size range further demonstrates the 
transferability of the nonorthogonal tight-binding approach. In Table IV, we have
also included the binding energies from Menon's NTB model for Ge$_n$ clusters
\cite{13}. Although the geometries of Ge$_n$ found in their work is almost the 
same as our results, the binding energies of Ge$_n$ starting from Ge$_5$ in Ref.13 
are about 0.10 $\sim$ 0.2 eV higher than our results and experimental values.

In summary, a nonorthogonal tight-binding model for germanium has been developed in 
this work. The transferability of the model is tested by various of bulk phases.
The agreements between NTB model and {\em ab initio} results for bulk solids and small 
clusters are satisfactory. For most Ge$_n$ cluster with $n=3$ to 10, the ground state 
geometries from tight-binding model coincide with those from {\em ab initio} calculation. 
The only exceptional cases are Ge$_8$ and Ge$_9$, in which the {\em ab initio} metastable
isomers are predicted as ground states by NTB scheme. However, the energy difference 
between the ground state configuration and the isomer is less than 0.01 eV/atom and 
within the accuracy of tight-binding approach. Therefore, the NTB model developed in 
this work can be applied to explore the configuration space of larger germanium clusters 
with $n>10$, for which a global minimization at the {\em ab initio} level is significantly 
more expensive. 
Our further studies shall include a genetic algorithm for sampling the phase space and 
finding possible low energy structural isomers of germanium clusters. Thus, first principle 
structural optimization can be performed on these local minima structures. On the other hand, 
this model will be also employed to simulate the thermodynamic properties such as melting 
and growth process of germanium cluster, which require a long time scale in TBMD simulation.

This work is partially supported by the U.S. Army Research Office
(Grant DAAG55-98-1-0298) and the National Natural Science
Foundation of China. The author (J.Zhao) are deeply grateful to 
Prof.E.Tosatti, Dr.J.Kohanoff, Dr.A.Buldum, and Prof.J.P.Lu for discussions.

\ \\
$^*$Corresponding author: zhaoj@physics.unc.edu


\begin{references}

\bibitem{1} C.Z.Wang, K.M.Ho, in {\em Advances in Chemical
Physics}, Vol.XCIII, p.651, Edited by I.Prigogine, S.A.Rice, (John Wiley $\&$
Sones, Inc., New York, 1996).

\bibitem{2} C.M.Goringe, D.R.Bowler, E.Herhandez, Rep.Prog.Phys.{\bf 60},
1447(1997).

\bibitem{3} Computational Material Science, Vol 12, No.3 (1998):
special issue on tight-binding molecular dynamics, Edited by L. Colombo.

\bibitem{4} M.P.Aleen, D.J.Tidesley, {\em Computer Simulation of Liqiuds},
(Clarendon Press, Oxford, 1987).

\bibitem{5} M.C.Payne, M.T.Teter, D.C.Allen, T.A.Arias, J.D.Joannopoulos,
Rev.Mod.Phys.{\bf 64}, 1045(1992).

\bibitem{6} G.Galli, A.Pasquarello, in {\em Computational Simulation in
Chemical Physicss}, edited by M.P.Allen and D.J.Tildesley, (Kluwer, Acedemic
Publisher, 1993), p.261.

\bibitem{7} L.Goodwin, A.J.Skinner, D.G.Pettifor, Europhys.Lett.{\bf 9},
701(1989).

\bibitem{8} C.H.Cu, C.Z.Wang, C.T.Chan, K.M.Ho, J.Phys.Condens.Matter {\bf 4},
6047(1992).

\bibitem{9} I.Kwon, R.Biswas, C.Z.Wang, K.M.Ho, C.M.Soukolis, Phys.Rev.B{\bf 49}, 
7242(1994).

\bibitem{10} M.Menon, K.R.Subbaswamy, M.Sawtarie, Phys.Rev.B{\bf 48}, 8398(1993).

\bibitem{11} M.Menon, K.R.Subbaswamy, Phys.Rev.B{\bf 50}, 11577(1994).

\bibitem{12} M.Menon, K.R.Subbaswamy, Phys.Rev.B{\bf 55}, 9231(1997).

\bibitem{13} M.Menon, J.Phys.Condens.Matter.{\bf 10}, 10991(1998).

\bibitem{14} M.van Schilfgaarde, W.A.Harrison, Phys.Rev.B{\bf 33}, 2653(1986).

\bibitem{15} J.C.Slater, G.F.Koster, Phys.Rev.{\bf 94}, 1498(1954).

\bibitem{16} C.Kittle, {\em Introduction to Solid State Physics}, (John Wiley $\&$
Sons, New York, 1986).

\bibitem{17} D.A.Papaconstantopoulos, {\em Handbook of the Band Structure of Elemental Solids},
(Plenum Press, New York, 1986).

\bibitem{18} W.A.Harrison, {\em Electronic Structure and the Properties of Solids},
(Freeman, San Francisco, 1980).

\bibitem{19} J.E.Kingcade, U.V.Choudary, K.A.Gingerich, Inorg.Chem.{\bf 18},
3094(1979); A.Kant, B.H.Strauss, J.Chem.Phys.{\bf 45}, 822(1966).

\bibitem{20} C.C.Arnold, C.Xu, G.R.Burton, D.M.Neumark, J.Chem.Phys.{\bf 102},
6982(1995).

\bibitem{21} P.W.Deutsch, L.A.Curtiss, J.P.Blaudeau, Chem.Phys.Lett.{\bf 270},
413(1997).

\bibitem{22} H.B.Huntington, in {\em Solid State Physics}, Vol.7, p.214, ed. by
F.Seitz and D.Turnbull, (Academic Press, New York, 1958).

\bibitem{23} J.J.Zhao, A.Buldum, J.P.Lu, C.Y.Fong, Phys.Rev.B{\bf 60}, 14177(1999).

\bibitem{24} D.Dai, K.Sumathi, K.Ralasubramanian, Chem.Phys.Lett.{\bf 193},
251(1992); D.Dai, K.Balasubramanian, J.Chem.Phys.{\bf 96}, 8345(1992);
D.Dai, K.Balasubramanian, J.Chem.Phys.{\bf 105}, 5901(1996).

\bibitem{25} J.R.Chelikowsky, S.Ogut, X.Jing, K.Wu, A.Stathopoulos, Y.Saad,
Mater.Res.Soc.Symp.Proc.{\bf 408}, 19(1996).

\bibitem{26} S.Ogut, J.R.Chelikowsky, Phys.Rev.B{\bf 55}, 4914(1997).

\bibitem{27} E.F.Archibong and A.St-Amant, J.Chem.Phys.{\bf 109}, 961(1998).

\bibitem{28} Z.Y.Lu, C.Z.Wang, K.M.Ho, Phys.Rev.B{\bf 61}, 2329(2000).

\bibitem{29} K.Raghavachari, C.M.Rohlfing, J.Chem.Phys.{\bf 89}, 2219(1988).

\bibitem{30} J.M.Hunter, J.L.Fye, M.F.Jarrold, Phys.Rev.Lett.{\bf 73},
2063(1993).

\end{references}
\end{document}